\documentclass[sigconf]{acmart}
\renewcommand\footnotetextcopyrightpermission[1]{}

\AtBeginDocument{%
  \providecommand\BibTeX{{%
    \normalfont B\kern-0.5em{\scshape i\kern-0.25em b}\kern-0.8em\TeX}}}

\usepackage[english]{babel}
\usepackage{graphicx}
\usepackage{graphics}
\usepackage[utf8]{inputenc}
\usepackage{tabularx}
\usepackage{makecell}
\usepackage{enumitem}
\usepackage{array,multirow}
\usepackage[final]{listings}
\usepackage{color}
\usepackage{subfig}
\usepackage{dblfloatfix} 
\usepackage{etoolbox}
\usepackage{float}
\usepackage{microtype}
\usepackage{blindtext}
\usepackage{xcolor}
\usepackage{filecontents}
\usepackage{booktabs}
\usepackage{siunitx}
\usepackage{pifont}
\usepackage{varioref}

\usepackage{xurl}
\usepackage{xspace}
\usepackage{hyperref}
\hypersetup{breaklinks=true}

\newcommand{\smartparagraph}[1]{\noindent{\bf #1}\ }

\begin{document}

\title{Characterizing Performance Inequity Across U.S. Ookla Speedtest Users}

\author{Udit Paul}
\affiliation{%
  \department{Department of Computer Science}
  \institution{UC Santa Barbara}
  \country{}}
  \email{u_paul@ucsb.edu}
 
\author{Jiamo Liu}
\affiliation{%
  \department{Department of Computer Science}
  \institution{UC Santa Barbara}
  \country{}}
  \email{jiamoliu@ucsb.edu}
 
  
\author{Vivek Adarsh}
\affiliation{%
  \department{Department of Computer Science}
  \institution{UC Santa Barbara}
  \country{}}
  \email{vivek@ucsb.edu}
  
\author{Mengyang Gu}
\affiliation{%
  \department{Department of Statistics and Applied Probability}
  \institution{UC Santa Barbara}
  \country{}}
  \email{mengyang@pstat.ucsb.edu}
  
\author{Arpit Gupta}
\affiliation{%
  \department{Department of Computer Science}
  \institution{UC Santa Barbara}
  \country{}}
  \email{arpitgupta@cs.ucsb.edu}
  
\author{Elizabeth Belding}
\affiliation{%
  \department{Department of Computer Science}
  \institution{UC Santa Barbara}
  \country{}} 
    \email{ebelding@cs.ucsb.edu}
\pagestyle{plain}
\pagenumbering{gobble}
\begin{abstract}

The Internet has become indispensable to daily activities, such as work, education and health care. Many of these activities require Internet access data rates that support real-time video conferencing. However, digital inequality persists across the United States, not only in who has access but in the quality of that access. Speedtest by Ookla allows users to run network diagnostic tests to better understand the current performance of their network. In this work, we leverage an Internet performance dataset from Ookla, together with an ESRI demographic dataset, to conduct a comprehensive analysis that characterizes performance differences between Speedtest users across the U.S. Our analysis shows that median download speeds for Speedtest users can differ by over $150$ Mbps between states. Further, there are important distinctions between user categories.  For instance, all but one state showed statistically significant differences in performance between Speedtest users in urban and rural areas. The difference also exists in urban areas between high  and low income users in $27$ states. Our analysis  reveals that states that demonstrate this disparity in Speedtest results are geographically bigger, more populous and have a wider dispersion of median household income. We conclude by highlighting several challenges to the complex problem space of digital inequality characterization and provide recommendations for furthering research on this topic.
\vspace{-10pt}
\end{abstract}
\maketitle

\section{Introduction}
The terms ``Internet inequity'' or ``digital inequality" refer to the gap in Internet access, access quality, and affordability that exists within and between geographic areas and communities or individuals of varying demographic attributes~\cite{PewDigitalDivide}. While the problem of Internet inequity in the U.S. has long existed~\cite{microsoft-access}, the Covid-19 pandemic has intensified its impact~\cite{PewInternet}. 
The lack of high quality, affordable Internet access severely impacts the outcomes of remote education, work from home, and telehealth, among others~\cite{deutche:economy,wallstreet:education,gtech:health,NBCRural,USAToday,TMF,Bloomberg}. 
Addressing this problem is critical; however, it is important  to first more deeply understand the challenges so that the right solutions can be applied to those communities most in need.  

A full characterization of Internet inequity requires combining  Internet access and quality data, at fine-grained geographic resolution, with  demographic datasets. Ideally, this data should be available at the granularity of census blocks (smallest demographic unit), or even smaller. While the Federal Communications Commission (FCC) and U.S. Census Bureau release related information at this granularity, the quality of this publicly available information is low. Through Form 477~\cite{fcc_477}, the FCC documents Internet coverage and theoretical maximum download speed from different Internet service providers at the granularity of census blocks. However, this dataset is known to inaccurately report and overstate Internet coverage, particularly in rural areas~\cite{broadband:report, bat}. More importantly, this dataset does not report the actual Internet performance experienced by the end users. This information is critical to characterize the regional quality of the Internet service.

Recently, Ookla~\cite{ookla_main} released an aggregated Internet performance dataset of Speedtest by Ookla measurements through the Open Data Initiative. This dataset overcomes a major limitation of Form 477 because it measures the Internet performance experienced by the end users at much finer spatial granularity. Additionally, Speedtest by Ookla is a popular Internet quality assessment solution, thereby facilitating a more fine-grained characterization of Internet inequity amongst its users in the U.S. While the scope of this dataset is limited to people who opt to take a Speedtest, it remains one of the largest end-user Internet performance measurement datasets, with high spatial fidelity, that is openly available to the public.  

In this work, we combine this Ookla dataset with geographic information from the U.S. Census Bureau and  demographic information provided by the Economic and Social Research Institute (ESRI)~\cite{ESRI-updated} to explore  multiple dimensions of Internet inequity in the U.S. amongst Speedtest takers. Our analysis shows the median download speed between two states in the U.S. could differ by as much as $150$~Mbps. We employ statistical techniques to quantify the extent of digital inequality between populations of different geographic locations (urban/rural) and income demographic variables (high/low income) within a state. Prior studies~\cite{census_digital_divide, ntia_digital_divide,pew_tech_adoption} undertook similar analysis using user survey data collected at the coarser geographic levels of census tract and county. Our analysis, however is conducted at the finer geographic granularity of census block group using actual network performance data. Confirming findings of~\cite{census_digital_divide, ntia_digital_divide}, our analysis shows that, for more than $45$ states, the quality of Internet for Speedtest users in rural areas lags behind that of urban areas. Further, we observe and quantify this divide in access quality between populations of urban areas; our analysis reveals a statistically significant difference in Internet quality between low income and high income block groups in $27$ states.  States that exhibit this divide between the urban Speedtest user groups tend to be bigger, with greater population and higher dispersion of household income compared to other states. Our findings demonstrate and quantify inequality of Internet access for Speedtest users and highlight the need for  thorough analysis of Internet performance experienced by  different communities.

In summary, our contributions are as follows:

\begin{itemize}
    \item  We aggregate and analyze an 18-month Internet performance dataset from Ookla Speedtest users from the 50 U.S. states, ESRI demographic data, and U.S. Census data to  identify and quantify key Internet performance inequities between  user groups in the U.S. based on geographic region and income.
    \item Using statistical techniques, we identify over 45 states where rural users receive statistically worse Internet performance than urban users over the 18-month period. 
    \item We further make novel observation and detect performance inequity between high income and low income urban Speedtest users in $27$ states.
    \item Through our analysis, we identify potential sources of bias in crowdsourced internet performance datasets such as Speedtest data.  Based on these findings, we conclude with specific recommendations for furthering research on Internet access inequality.
\end{itemize}



\section{Datasets}
\label{sec_data_tools}

In this section, we describe the publicly available datasets that we use for our analysis.  Our analysis is based on data available throughout 2020 and the first half of 2021.  This data is aggregated at the time-granularity of quarters by Ookla, as described below.

\vspace{-8pt}
\subsection{Performance Data}
The quality of user experience for web-based activities is dictated by available upload and download speed and latency to the remote server. For example, the user experience for video streaming applications depends mostly on available download speed, video conferencing applications depends on both the upload and download speeds, and web browsing depends mostly on latency. The Ookla Speedtest allows users to assess the quality of their Internet connection using either the web-based portal or native mobile application~\cite{ookla_main}. Ookla relies on volunteer users to conduct a speed test that measures Internet download and upload speed and latency at the current connection point. For each request, Ookla's controller uses the client's location to select a set of measurement servers that are geographically closest to the client. It then chooses the one with minimum round-trip time (RTT) as an endpoint for the test.  Ookla dynamically scales each measurement with multiple parallel connections to saturate the bottleneck link. To ensure high quality and fidelity data is obtained, Ookla operates a network of tens of thousands of measurement servers, and periodically eliminates servers that perform poorly~\cite{ookla-servers}.  The results of a single test offer an instantaneous snapshot of Internet performance at the current location, to the current point of attachment, and subject to the current competing traffic on the path to the selected server.  Together, the aggregation of many of these measurements can paint a picture of connectivity within a given geographic area that is broadly diverse both in time, exact physical location, and network traffic load. While access quality can change greatly within a small spatial area based on subscribed plan, residential vs. business connectivity, etc., measurement aggregation can still offer broad insight into general performance trends for a region, as we will demonstrate through our analysis.

Through the Open Data Initiative, Ookla has released an aggregated version of the data it collects to the public every quarter, beginning in January 2019~\cite{ook_open_data}. In this dataset,  geographic areas are grouped into quad tiles~\cite{ookla-open-github}. The size of these quad tiles depends on their geographic location. For example, quad tiles measure approximately $600$ by $600$ sq. meters at the equator, and roughly $500$ by $500$ sq. meters in Los Angeles. The dataset reports the average of all measurement values for each quarter of the year for each tile. Ookla divides the data into two groups, each with measurements from users connected to the Internet via a (1)~fixed broadband network (e.g., Cox, Xfinity, etc.);  and (2)~mobile network (e.g., T-Mobile, US Cellular, etc.). Ookla only includes the measurements from mobile devices with in-built GPS. This filter ensures higher accuracy in mapping Speedtest measurements to geographic locations (or tiles).

Given our focus on the U.S., we filter the tiles from Ookla's dataset to include only those that are completely within the geographic boundary of the U.S. The total number of tiles with measurements of fixed networks are $1.70$ million (M), $1.74$ M, $1.72$ M and $1.51$ M in each of the respective four quarters of 2020. In the first and second quarter of 2021, approximately $1.53$ M tiles are present in the dataset. The number of tiles in the mobile network group was approximately $600$~k in each quarter of 2020 and 2021. 
The number of measurements from each tile depends on multiple factors, such as population density, popularity of the speed test application, etc. Because the number of tiles for the mobile network group is very small, we focus our analysis on the fixed broadband network.

\smartparagraph{Critique.} The potential shortcomings of crowdsourced Internet measurements using tools such as Ookla's Speedtest are studied in the literature~\cite{feamster_measure,igor_ookla}.  Essentially, these crowdsourced measurements may introduce bias in terms of locations where the tests originate and network conditions under which they are conducted. However, in the absence of true underlying distributions, the effect and magnitude of this bias is  difficult to quantify. 

\begin{figure*}[t]
  \includegraphics[width = 0.9\linewidth]{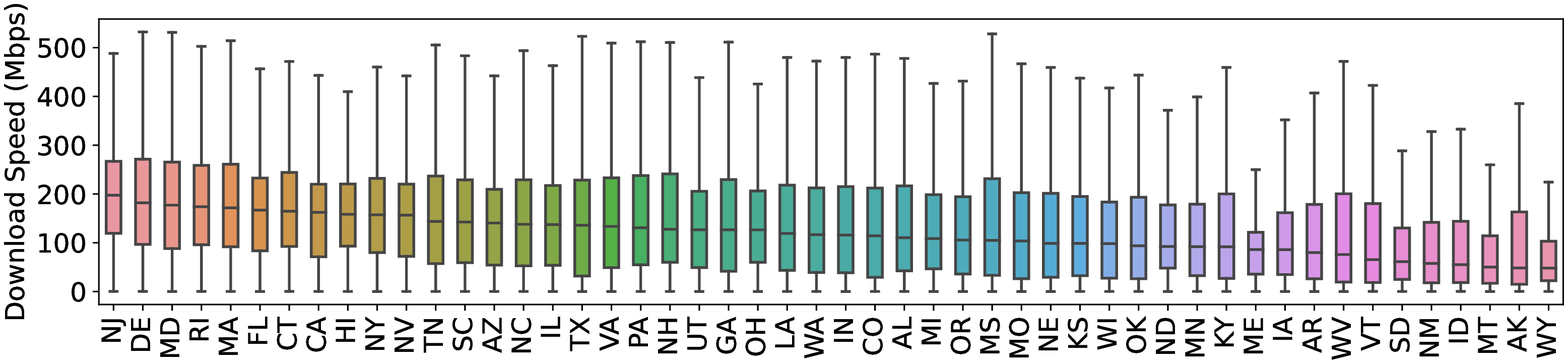}
  \vspace{-8pt}
  \caption{Distribution of download speed of each state during Q2-2021.}
  \label{fig:q6_speed_states}
  \vspace{-15pt}
\end{figure*}

While Ookla does not have any research data on the specific demographic attributes of their user population, they do have general information about Speedtest usage.  According to their data, people tend to use Speedtest in a variety of circumstances, including when they are having Internet issues, immediately after setting up a new device, and when they arrive at a new location (i.e. hotel, public space, or even other side of the house).  With the shift to education and work-from-home due to the Covid-19 pandemic, an increased number of users tested home Internet connections, to discern both the number and types of applications that could be concurrently supported, as well as the locations in the house that offered the best connectivity.  Hence, while it is impossible to say that Ookla data does not have bias towards certain types of events or points of attachment, the aggregated data, grouped over both space and time, offers a broad swath of usage scenarios.  Our goal is to study the network performance, as represented by Ookla Speedtest results, during those scenarios, and to attempt to correlate performance with demographic data to the extent possible.  Because of potential bias, we cannot definitively characterize Internet connectivity in a given quad tile; however, we hope that our work is a step forward in that direction and can point to where additional data and analysis is needed.

\subsection{Demographic Data}
To study the relationship between Internet performance, geographic region, and user demographic attributes, we leverage the demography data provided by ESRI's Updated Demographics~\cite{ESRI-updated}. ESRI curates this 
dataset using multiple sources that provide current-year estimates and 5-year projections of a variety of demographic attributes. This dataset is a critical combination of the most recent demography data available that is also highly accurate~\cite{esri-accuracy}.

Most demographic information is aggregated and released at the granularity of  census block groups~\cite{acs5yeardata}. For our analysis, we choose the demographic attribute of median household income; prior work~\cite{deutche:economy,wallstreet:education} has shown income to play an important role in Internet access availability to different user groups. Using the ESRI dataset~\cite{esri-dataset}, we obtain the median household income at the granularity of the census block group in the U.S.  At this granularity, the ESRI dataset is comprehensive and covers $98.6\%$ ($214$K out of $217$K) of all  block groups in the U.S. In addition to median household income, we also obtain the population of each census block group in the U.S. using~\cite{esri-dataset}.

To the median household income data, we also include the type of geographic area (urban/rural); again, prior studies have shown that the region type has an impact on Internet access availability and quality~\cite{bat}. In contrast to median household income, the distinction between rural and urban area types are  made at the level of census blocks. On average, there are $39$ blocks present in a census block group~\cite{wiki_block}. We utilize data from the 2010 U.S. Census \cite{2010TigerLineShapefile} to first obtain each census block's designation. Subsequently, we aggregate all census blocks within a census block group to classify the area type of the census block group based of the area type of a simple majority of the blocks within that group. Our aggregation allocates $22.7\%$ of the total U.S. population of $330$M to rural block groups. This percentage of rural population is consistent with the number reported by the Census~\cite{census_population, wiki_population}.

\begin{table*}[t]
    \vspace{-15pt}
    \resizebox{0.9\linewidth}{!}{%
  \begin{tabular}{l|cc|cc|cc|cc|cc|cc}
    \toprule
    \multirow{2}{*}{Metric} &
		\multicolumn{2}{c|}{Q1-2020} &
		\multicolumn{2}{c|}{Q2-2020} &
		\multicolumn{2}{c|}{Q3-2020} &
		\multicolumn{2}{c|}{Q4-2020} &
		\multicolumn{2}{c|}{Q1-2021} &
		\multicolumn{2}{c}{Q2-2021} \\
		& {Median} & {IQR} & {Median} & {IQR} & {Median} & {IQR} & {Median} & {IQR} & {Median} & {IQR} & {Median} & {IQR}\\
		\midrule
		Download (Mbps) & 86.49 & 121.65 & 86.46 & 123.98 & 92.18 & 131.72  & 105.84 & 144.80 & 114.12 & 152.80 & 126.34 & 173.73\\
		Upload (Mbps) & 11.68 & 18.54 & 11.74 & 19.04 & 12.25 & 21.99 & 13.71 & 24.92 & 14.77 & 27.50 & 15.37 & 29.19 \\
		Latency (ms) & 18 & 17 & 18 & 16 & 18 & 16 & 16 & 13 & 16 & 13 & 16 & 13 \\
    \bottomrule
  \end{tabular}
  }
  \caption{Aggregated statistics of all network metrics across in 2020 and 2021, by quarter.}
  \label{tbl_agg_stats_ookla}
\vspace{-22pt}
\end{table*}


\subsection{Curating an Aggregate Dataset}
To understand the effect of different location types and demographic attributes on Internet performance, we first need to assign each tile in the Ookla dataset into the much larger areas of census block groups because demographic information is only available at the granularity of block group. Because the Ookla data provides the geographic coordinates of each tile, we are able to allocate tiles to the polygon boundaries of the census block groups. Post allocation, analysis shows these tiles are present in roughly $94\%$ of all census block groups in the U.S. in each quarter of 2020 and the first two quarters of 2021. To determine the Internet performance of a particular block group, we take the average of each network metric for all the tiles that belong to that block group, and we weight by the number of tests that originated from each tile. With this level of aggregation, we are able to quantify the Internet performance of census block groups within a state, placing higher weights to tiles that have more number of tests.

\subsection{Aggregated Dataset Overview}
Prior work~\cite{anjaIMC20, liuFeamsterPAM21} has demonstrated that, due to the  Covid-19 pandemic, Internet traffic patterns changed significantly through 2020; traffic volume increased, and significant capacity upgrades were made by service providers to meet the rising demand. This change in Internet traffic dynamics is also captured in our high level analysis of our aggregated dataset. 

\begin{figure}[t]
  \includegraphics[width=0.7\linewidth]{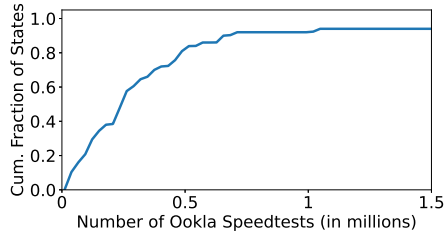}
  \vspace{-10pt}
  \caption{CDF of Ookla Speedtests across states in the U.S. in Q2-2021.}
  \label{fig_test_state}
  \vspace{-15pt}
\end{figure}


Table~\ref{tbl_agg_stats_ookla} presents the median download speed, upload speed and latency (along with the inter-quartile range (IQR)) recorded across all the tiles within the boundary of the U.S. over the six quarters. Both the median download speed and upload speed increase during our studied time period. 
The recorded median latency also improves, decreaseing from $18$ms from the first three quarters of 2020 and remaining at $16$ms from Q4-2020 to Q2-2021.

Figure~\ref{fig:q6_speed_states} shows the distribution of download speed in every state in the U.S. in Q2-2021. New Jersey, Delaware, Maryland, Rhode Island and Massachusetts recorded the best median download speeds during Q2-2021. During the same quarter, Wyoming emerged as with the lowest download speed along with Arkansas, Montana, Idaho and New Mexico. The median difference in download speed is $149.40$ Mbps between New Jersey and Wyoming during Q2-2021. The median download speed of New Jersey and Delaware  remained within the top five during Q1-2021 and all of 2020. Wyoming and New Mexico, on the other hand, recorded the lowest median download speed our studied time period. 

New Jersey, Delaware, Rhode Island and Maryland also had the highest median upload speeds during Q2-2021, while New Mexico and Wyoming continued to have the lowest. The best and worst performing states by median upload speed differed by $27.63$Mbps in Q2-2021. As observed in the case of download speed, the best and worst performing states remained consistent during other the quarters of 2020 and 2021. Similar trends were  observed for latency.

Another critical difference in state-by-state data is the number of tests that originate from each state as shown in Figure~\ref{fig_test_state}. There were $2.8$M  Ookla Speedtests conducted in California in Q2-2021, and $1.9$M and $1.6$M tests conducted in Texas and Florida, respectively, representing the greatest number of tests by volume in the nation. Once the number of tests are normalized by the population of the respective states, there are $0.07$ tests conducted per person (pp) from each of these states. These same three states also recorded the greatest number of tests during 2020 and Q1-2021. North Dakota had the fewest  tests ($25$K or $0.03$ pp) followed by South Dakota ($26$K or $0.03$ pp) and Alaska ($34$K or $0.04$ pp) in Q2-2021. These states also produced the fewest tests in the other quarters analysed in this work.


\section{Methodology: Quantifying Impact on Performance}
\label{sec_method}


To understand the relationship between Internet performance, region type, and median income within a state, we need to compare the Internet performance received by different user groups in the Speedtest dataset. Ideally, controlled experiments would be conducted to explore this relationship. However, as pointed out in~\cite{Bischof_need}, Speedtest users are not randomly selected, thereby rendering controlled experiments infeasible in this scenario. To overcome this limitation, similar to the methodology followed in the diverse fields of epidemiology, sociology and economics, we utilize natural experiments to conduct our analysis. By employing natural experiments, we are able to pair two user groups who both conducted Speedtests while differing in terms of the location or median household income. This pairing imitates randomness and allows us the opportunity to explore the relationship between these factors and Internet performance of Speedtest user groups~\cite{Bischof_need, natural_exp1}.

For every factor we consider in our natural experiments, we set a null hypothesis, $H_0$, pose a hypothesis ($H$) and compare the average Internet performance of two different types of block groups in a state. These block groups can differ in terms of location type (urban/rural) or median household income (high/low income). This approach allows us identify Internet inequity in the dimensions of location and income within every U.S. state. Table~\ref{tbl_hypothesis} presents the $H_0$ and $H$ for the factors of location and income.

\begin{table} [!ht]
\centering
\begin{tabular}{| m{28pt} | m{75pt} |  m{90pt}|}\hline
Factor& \quad\quad\quad $H_0$ & \quad\quad\quad\quad $H$\\\hline
Location & No difference in urban and rural Internet performance & Urban Internet performance is better than rural Internet performance\\\hline
Income & No difference in urban high income and urban low income internet performance & Urban high income Internet performance is better than urban low income Internet performance \\\hline
\end{tabular}
\caption{Null and alternative hypotheses in natural experiments for detecting statistically significant difference in performance between user groups.}
 \label{tbl_hypothesis}
 \vspace{-27pt}
\end{table}

To compare and detect a statistically significant difference in Internet performance between Speedtest user groups, we employ two methodologies: i) one-tailed Komolgorov-Smirnov (K-S) 2~sample test~\cite{ks2sample} and ii) one-sided Mann–Whitney U (M-W U) test~\cite{mwutest}. We employ these two separate tests to reduce the number of false positives in detecting statistically significant differences in performance between two user groups. These tests are non-parametric and, therefore, can be used for non-normal distributions. The K-S test captures the difference between two samples by evaluating the maximum distance between the two distributions for a confidence interval, $\alpha$. However, the K-S test is less sensitive to the difference in median between two distributions. The M-W U test, on the other hand, detects the discrepancy between the mean ranks of the two groups being compared and hence is more capable of detecting a change in median values between the groups. To further ascertain statistical significance and reduce Type-I errors~\cite{type_errors} while employing these two tests, we employ the Bonferroni correction technique~\cite{bonferroni} for multiple testing.

For the K-S test, we only consider two distributions to be statistically different when the $p-value$ is below $0.05$, and the test statistic (also known as $D-statistic$) is greater than the corrected threshold value for each experiment. Similarly, in the case of the M-W U test, we consider a test statistic significant if the $p-value$ is below $0.05$. For each test we conduct between two distributions, 
we analyze whether there is a presence of strict/strong or weak conformance to our expected hypothesis. We consider two distributions to be strictly different if one distribution remains statistically different from the other over the entire distribution. Suppose a distribution leads another for some part while lagging otherwise, both with statistical significance. In that case, we consider such a crossover case to be weak. The one-tailed nature of the two tests allows for the evaluation of $test-statistic$ ($D$ in the case of K-S test) and $p-value$ ($p$ for both the K-S test and M-W U test) for both our hypothesis and the null hypothesis. By so doing, we are able to quantify the extent to which we failed to reject the null hypothesis and accept our hypothesis for each test we conduct.

\begin{figure*}[t]
    \centering
    \subfloat[Georgia]{\label{a}\includegraphics[width=.23\linewidth]{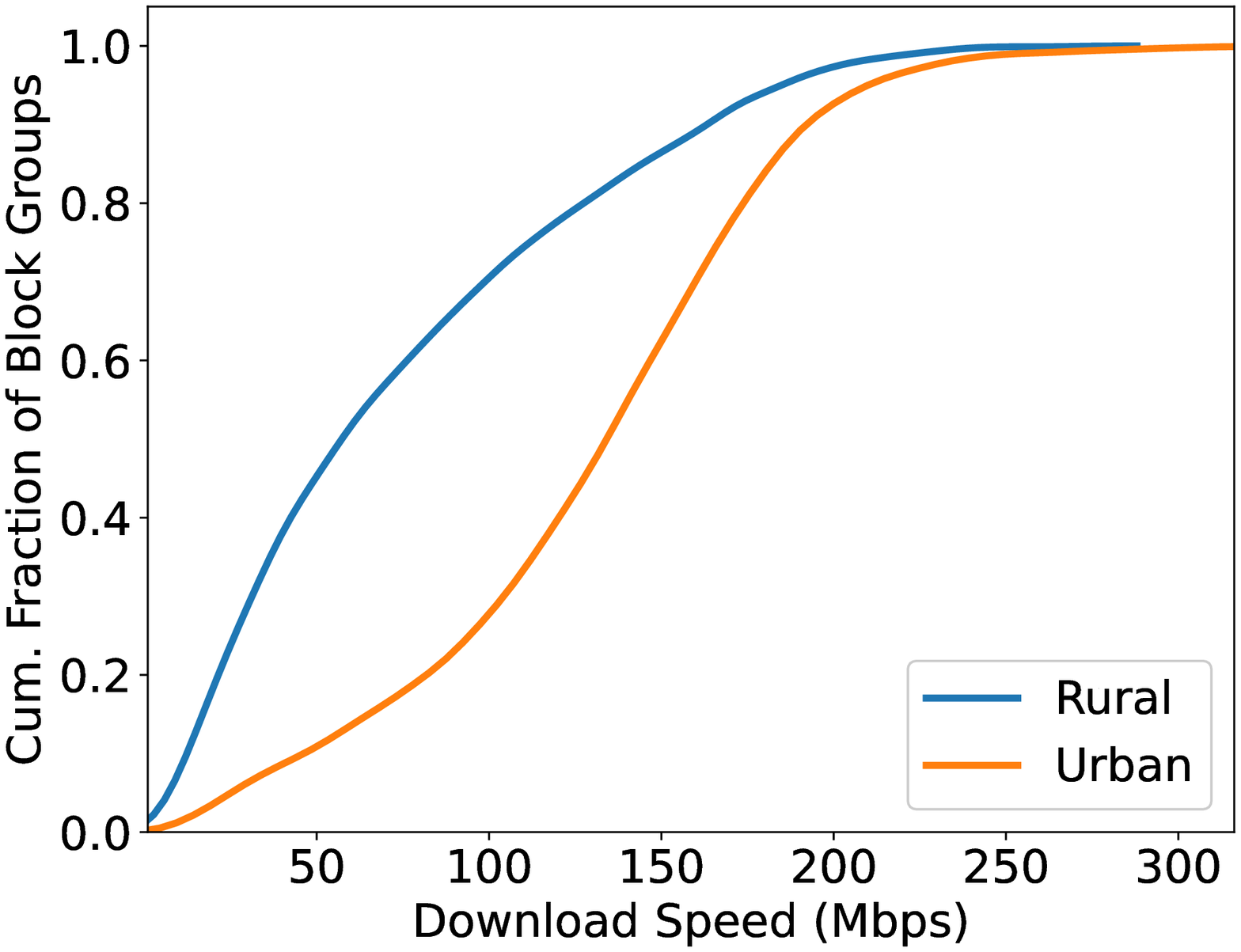}}\hfill
    \subfloat[Louisiana]{\label{b}\includegraphics[width=.23\linewidth]{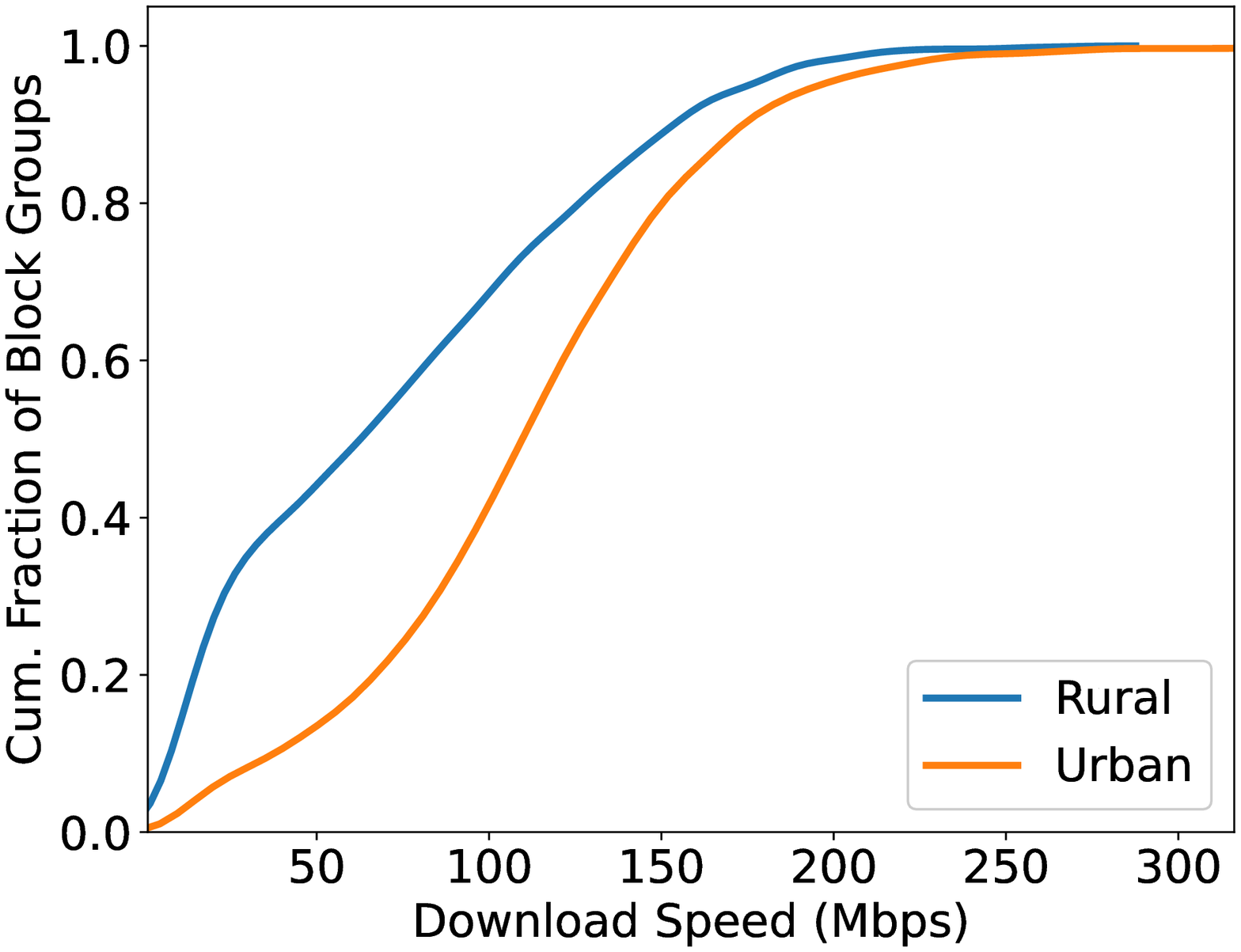}}\hfill
    \subfloat[Rhode Island]{\label{b}\includegraphics[width=.23\linewidth]{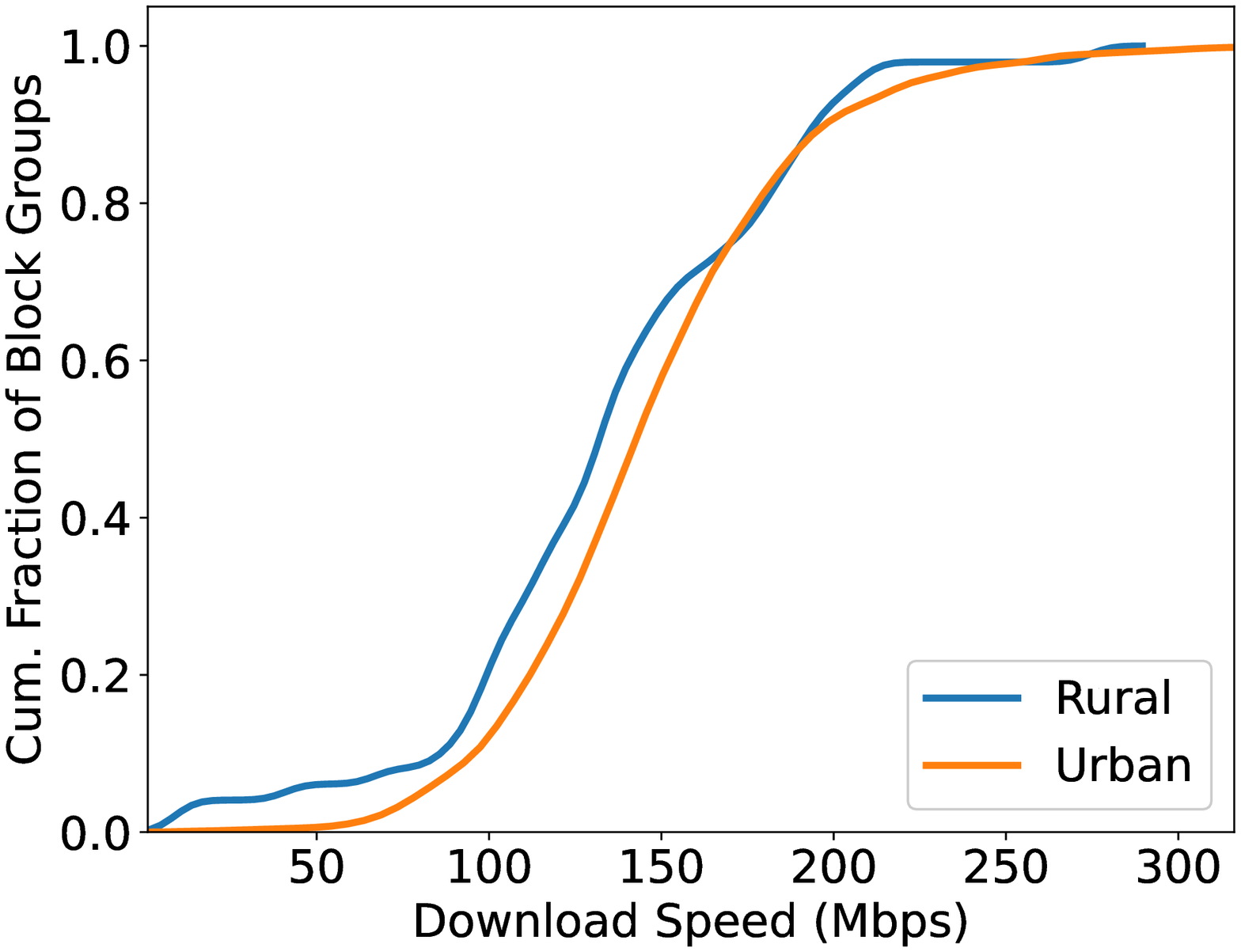}}\hfill
    \subfloat[Delaware]{\label{b}\includegraphics[width=.23\linewidth]{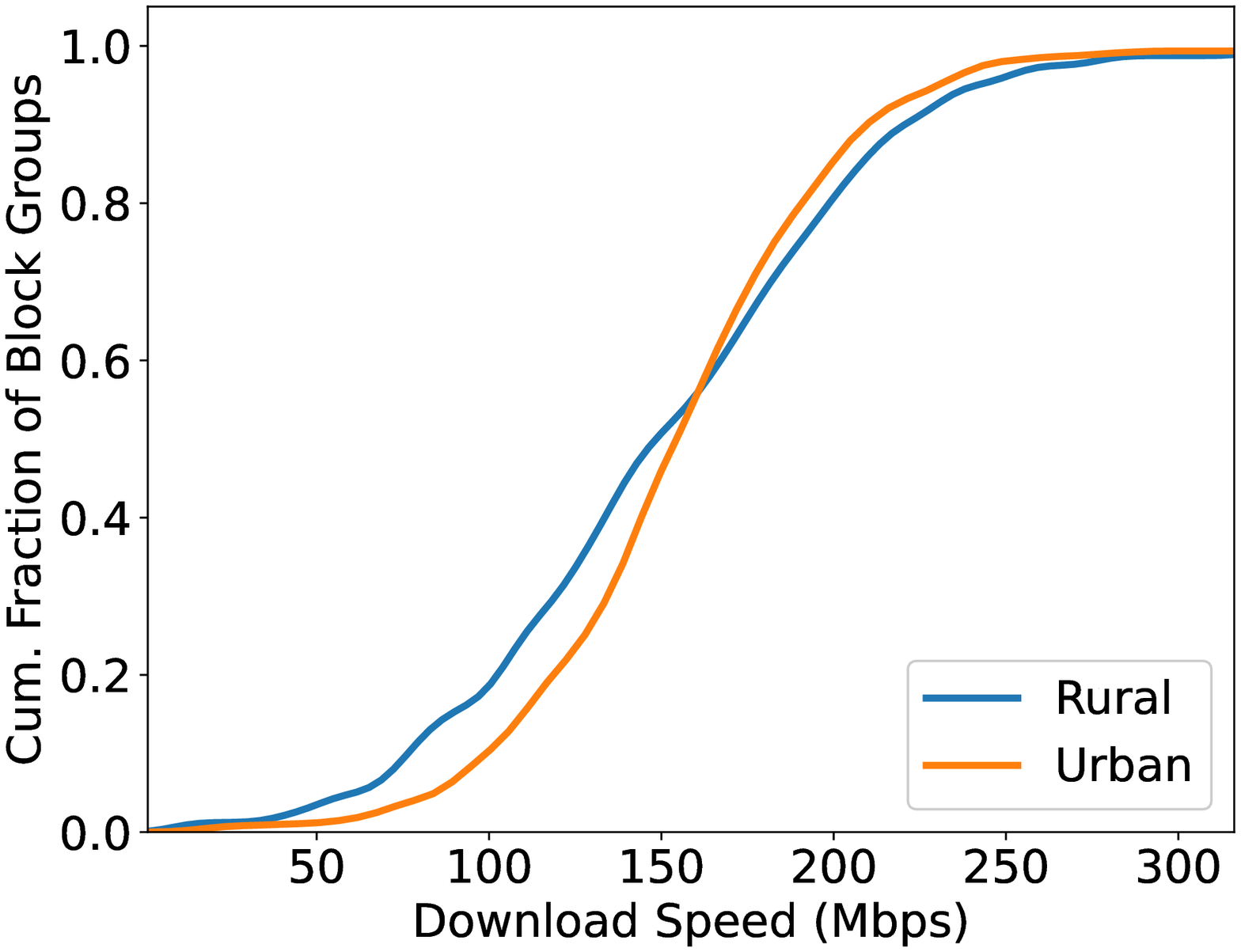}}\hfill
    \vspace{-8pt}
    \caption{Q1-2020 CDFs of download speeds, disaggregated by rural  and urban block groups, in four example states.}
    \label{fig:q1_speed}
\vspace{-10pt}
\end{figure*}


Suppose there is no statistical difference between the distributions of Internet performance of two different groups in a state. In that case, we conclude that the association between region type or demographic attribute and Internet performance metric is not statistically significant in that state. We employ these two tests across all the states in the country.

\section{Performance within a State}
\label{sec_factors}

In this section, we analyze how Internet performance varies between different population groups of Speedtest users within U.S. states using the methodology described in Section~\ref{sec_method}. We present our findings and highlight specific states where this performance difference is detected across location type (Section~\ref{sec_ur}) and median household income (Section~\ref{factor_income}).

\vspace*{-0.1in}
\subsection{Impact of Location Type}
\label{sec_ur}

We first employ our methodology to show the impact of location type on Internet performance. Using the datasets described in Section~\ref{sec_data_tools} and location hypotheses stated in Table~\ref{tbl_hypothesis}, we quantify the difference in performance between urban and rural block groups within every U.S. state. 

We  evaluate the block group level distributions of the network metrics for urban and rural areas across all states. For the K-S test, we compare the distribution for each state per network metric and either accept or reject  $H_0$ based on the $D$ and $p$ values obtained from each comparison. Similarly, for the M-W~U test, we accept or reject  $H_0$ based on the resulting $p$ value obtained by conducting the one-sided test. We conduct this analysis on every quarter to understand the change in performance over the course of 2020 and the first two quarters of 2021.

We begin our analysis with Q1-2020.  For both tests we observe a statistically significant difference (strict) in distributions of urban and rural block group download speeds in favor of our hypothesis, $H$, in $47$ states in the country. The states that did not show a difference were Rhode Island, Delaware, and Connecticut. Figures~\ref{fig:q1_speed}(a) and (b) present two examples of states (Gerogia (GA) and Louisiana (LA)) where the rural block groups recorded statistically poorer performance than their urban counterparts across the studied time period (CDFs of Q1-2020 are shown as examples). In the case of Rhode Island (RI) and  Delaware (DE), as illustrated in Figures~\ref{fig:q1_speed}(c) and (d), the performance in the urban block groups was not statistically better than that of rural block groups (for RI in Q1 and DE in both Q1 and Q4). Subsequent analysis of the remaining quarters using both tests revealed $48$, $49$ and $49$ states conforming to our hypothesis in Q2, Q3 and Q4 of 2020, respectively. The number of conforming states remained $49$ for the first two quarters of 2021.

We repeat the  analysis to detect statistically significant differences in upload speeds between rural and urban block groups. Results show the urban block groups within $47$ states outperform their rural counterparts in Q1-2020. Similar to the case of download speed, RI  did not show any difference in  upload speed between these two groups. However, unlike in the case of download speed, New Hampshire and North Dakota rural block groups outperformed urban block groups. Q3-2020 recorded $49$ conforming states. Q2 and Q4 of 2020, as well Q1- and Q2-2021, recorded $47$ states that strictly conformed with our hypothesis. When analyzing latency, $47$ and $46$ states conformed to our hypothesis in Q1- and Q2-2020, with the number reducing to $45$ in both Q3- and Q4-2020. The number of conforming states increased to $46$ and $47$ for Q1- and Q2-2021, respectively.

We next evaluate our location hypothesis on the number of Ookla Speedtests per person that originate from the two types of block groups. We set  $H_0$ as the number of Speedtests per person originating from rural block groups is not less than the number of Speedtests per person conducted in  urban block groups. 
We pose the hypothesis ($H$) that the number of Ookla Speedtests that are conducted per person in urban block groups will be greater than that originating from rural block groups. About $20$ states conformed to this hypothesis in each quarter of 2020 and 2021. The remaining $30$ states did not demonstrate any statistically discernible difference in the normalized Speedtest counts between these two location types.

\noindent
{\bf
Takeaways.} 
Our analysis of Internet performance for  Speedtest users  shows a clear divide between rural and urban regions 
in nearly every state in the country. In total, $49$ different states conformed with our hypothesis and demonstrated a statistically significant difference (in favor of urban areas) in performance in at least one of the six quarters we analyzed. Rhode Island, the smallest state in the U.S. by land mass~\cite{ri_small_size}, remained  the sole exception, where a statistically significant difference in performance was not found in any quarter. Its smaller size could translate into ease in establishing network infrastructure throughout the state, as opposed to  bigger states with larger, and in part more difficult, terrains. Further, while overall performance improved during the studied time period, the rural block groups continued to fare worse than the urban block groups in a vast majority of the states. However, the average K-S test $D-statistic$ value between urban and rural block groups across all the states reduced from $0.45$ in Q1-2020 to $0.40$ in Q2-2021, representing a small narrowing of the divide between these location types. 
Finally, and surprisingly, our  results  indicate that, once normalized by the total population, the majority of states do not exhibit a bias towards the urban areas in the number of tests per person.

\begin{figure*}[t]
\centering
\subfloat[Georgia]{\label{a}\includegraphics[width=.23\linewidth]{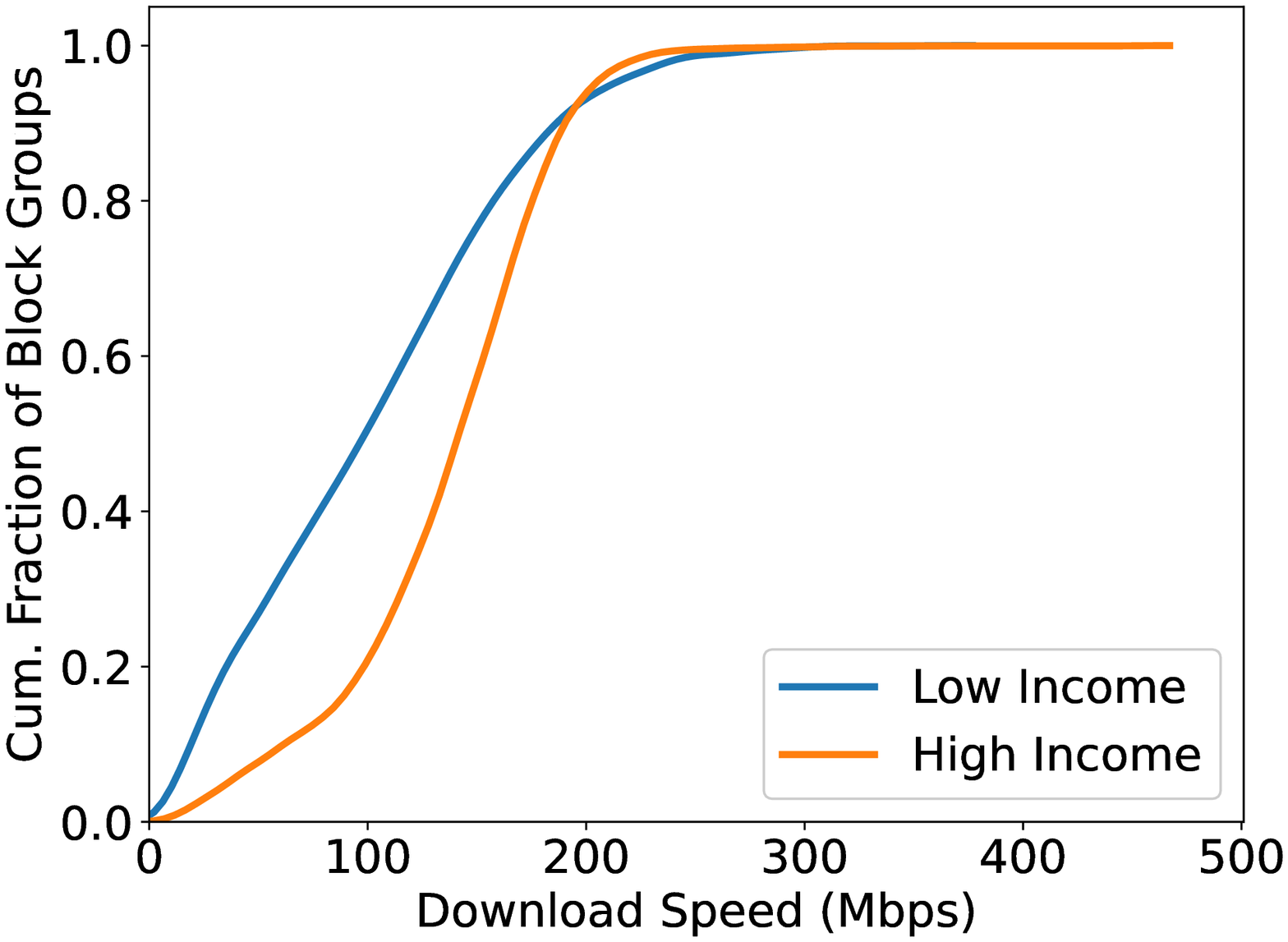}}\hfill
\subfloat[Louisiana]{\label{b}\includegraphics[width=.23\linewidth]{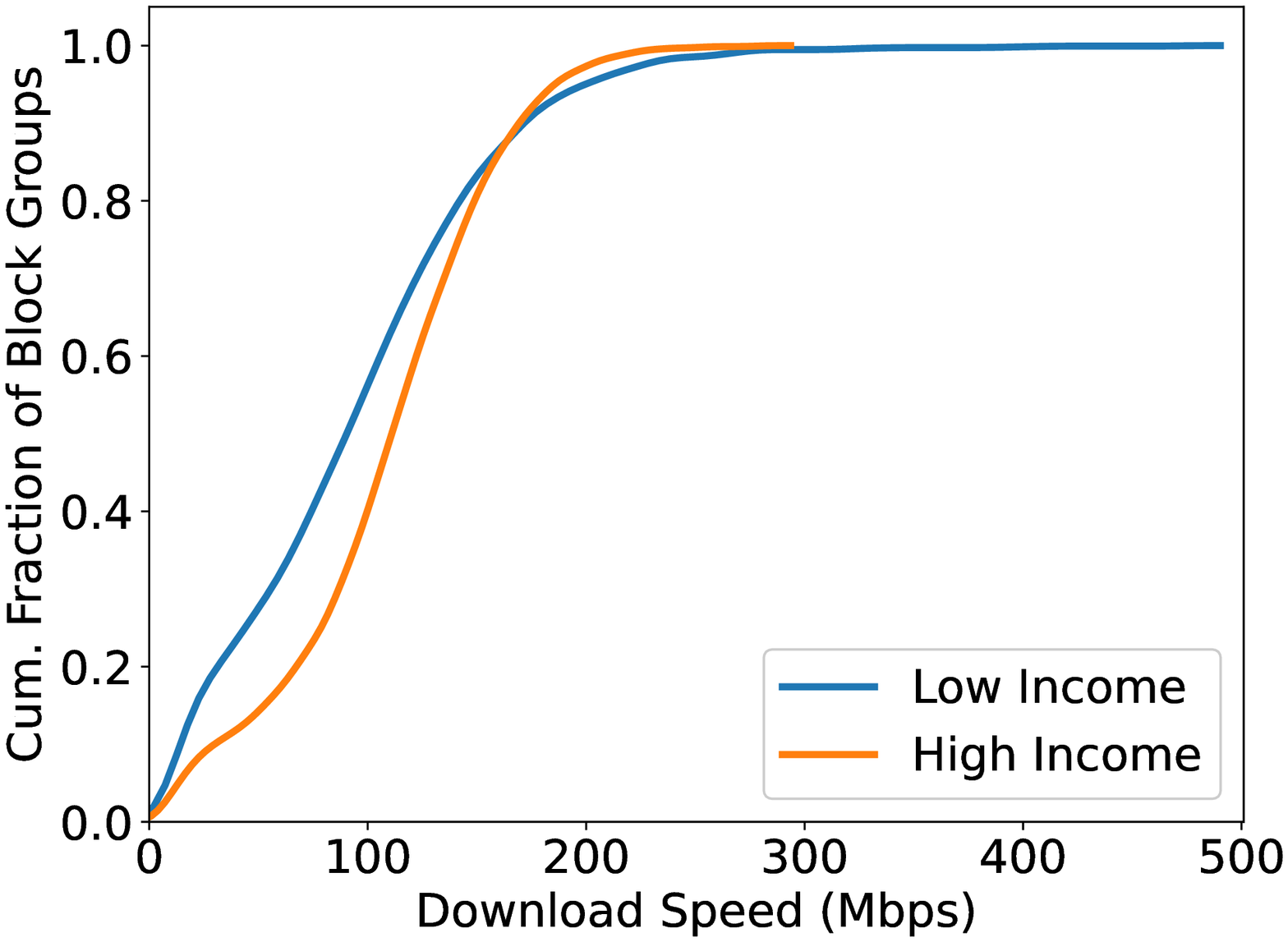}}\hfill
\subfloat[Rhode Island]{\label{c}\includegraphics[width=.23\linewidth]{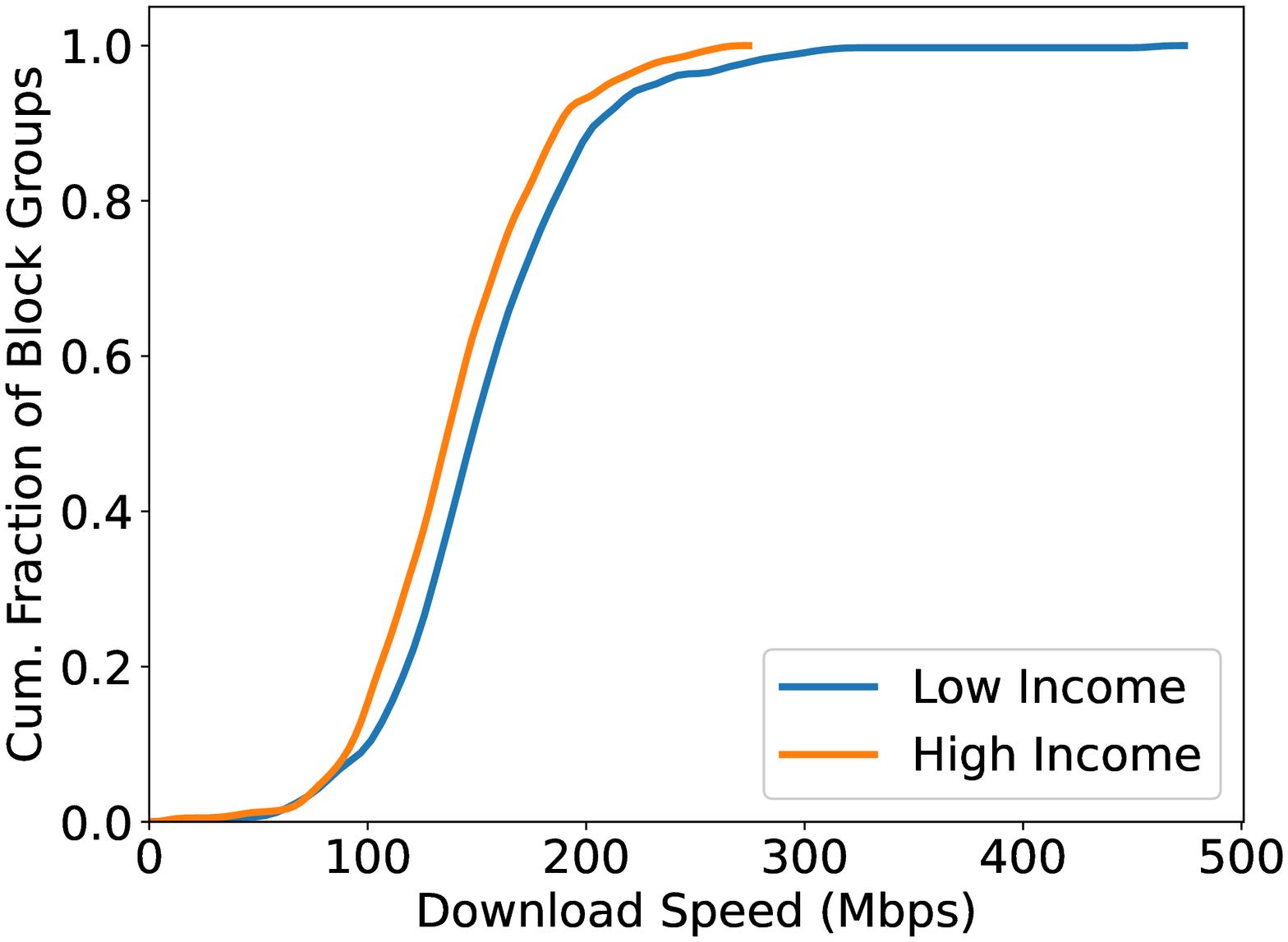}}\hfill
\subfloat[New Jersey]{\label{d}\includegraphics[width=.23\linewidth]{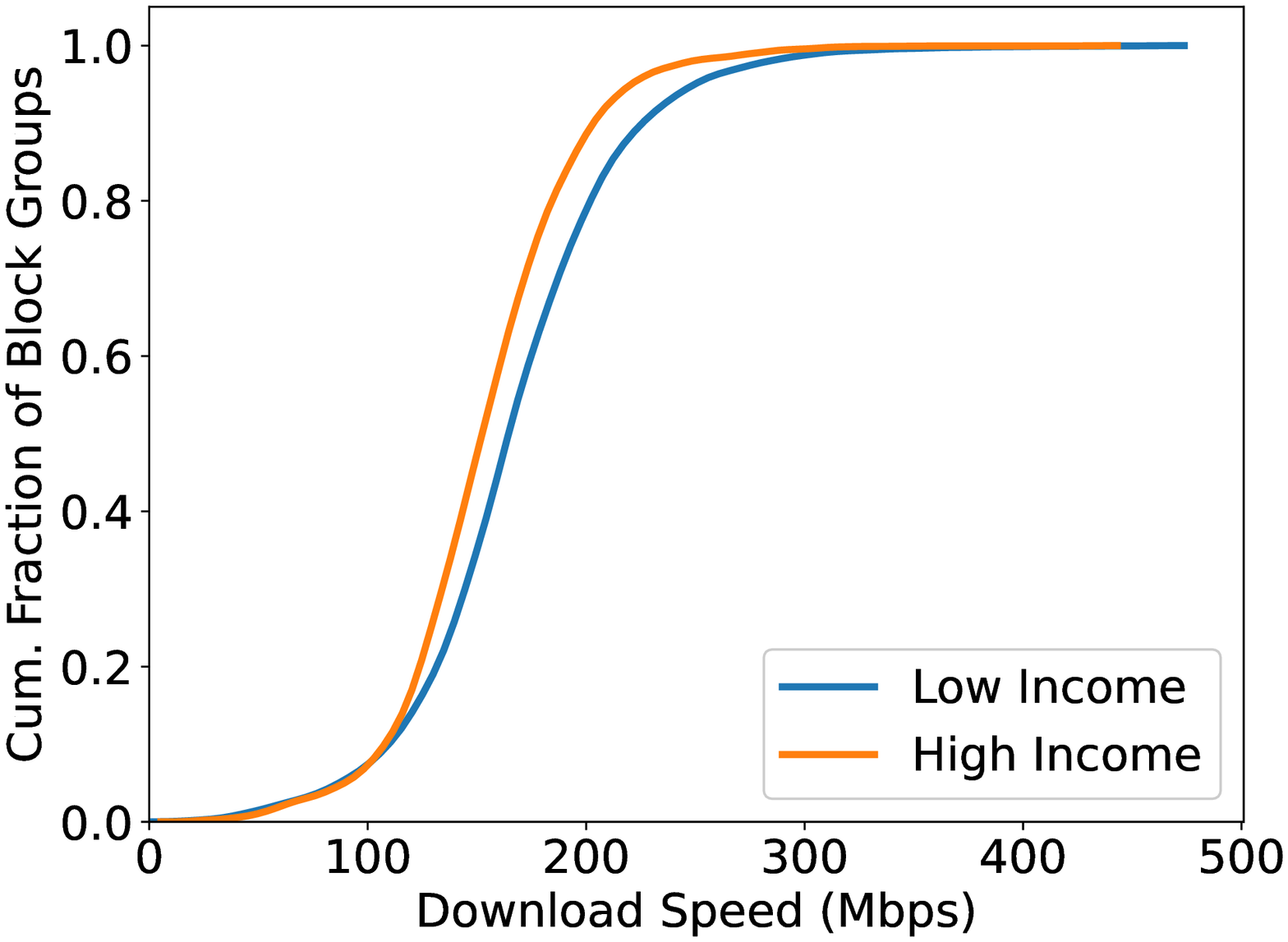}}\hfill
\vspace{-8pt}
\caption{Q1-2020 CDFs of download speeds, disaggregated by low  and high income block groups, in four example states.}
\label{fig_q1_speed_income}
\vspace{-12pt}
\end{figure*}

\subsection{Impact of Median Household Income}
\label{factor_income}
We next explore the relationship between an important demographic variable, median household income, and Internet access quality. Prior work~\cite{rural_income_same} indicates that rural areas tend to have higher poverty and lower median household income. Given the presence of this relationship between location type and income, we analyse the impact of median household income on Internet quality in urban census block groups. To do so, we divide the urban block groups within every state into two categories. Based on the median block group level household income obtained from the ESRI dataset, we calculate the average state household income for every state in our dataset. With this average state income, we classify the block groups whose median income lay below the state income as ``Low Income" block groups and those with income greater than or equal to the state income as ``High Income'' block groups. Our null hypothesis $H_0$ and alternative hypothesis $H$ for income are presented in Table~\ref{tbl_hypothesis}.



Similar to our location related analysis, we begin our exploration on the Ookla data from Q1-2020. Table~\ref{tbl_test_results} presents the number of states where a statistically significant difference in download speed was detected by the K-S test and M-W U test between high income and low income census block groups. Figures~\ref{fig_q1_speed_income}(a) and (b) present examples of states, Georgia and Louisiana, that strictly conform with our hypothesis in both tests across all quarters. Arkansas, Alabama, and New Mexico are examples of other states that demonstrated similar trends. The M-W U test detected an additional nine states with statistically significant differences in download speed compared to the K-S test. Texas, Alaska and California were amongst the nine states that only weakly passed the K-S test; i.e., for some download speed range, the low income neighborhoods outperform the high income neighborhoods. Examples of states where the higher income block groups recorded statistically worse download speed compared to the lower income block groups in both tests are RI and New Jersey (NJ) as observed from Figures~\ref{fig_q1_speed_income}(c) and (d). This observation is captured through the $D$ value of $0.14$ ($0.15$ in RI) with the $p$ value of $1.04\times10^{-23}$ ($1.84\times10^{-04}$ in RI) in favor of $H_0$ while using K-S test. Other states that exhibit a similar pattern include Delaware and Massachusetts. 

In the next two of quarters of 2020, the number of states that conformed with our hypothesis remained fairly similar to Q1-2020. Q4-2020 (as well as first two quarters of 2021) witnessed a rise in the number of conforming states for both tests. Across these six quarters, $14$ states conformed to our hypothesis in every quarter for K-S test. This number increases to $27$ states for the case of M-W U test. This indicates the presence of a large number of states where the median download speed of high income block groups remained statistically better than the low income block groups over the course of $18$ months.

We now turn to upload speed, in which the number of states where higher income block groups achieve better performance than lower income block groups is $31$ (K-S test) and $35$ (M-W U test) in Q1- and Q2-2020. Examples of states where our hypothesis is strictly held include Louisiana, Arkansas and Virginia. While there are a number of states where the low income block groups do not exhibit statistically worse upload speed than the high income block groups, the converse does not hold for any state. A similar trend is observed during the rest of 2020 and first two quarters of 2021. In the case of latency, the number of conforming states remained fairly similar for the K-S test. The M-W U test, however, had more conforming states than to the K-S test except for Q2-2021.

Finally, we investigate whether there exists a statistical difference in the number of Ookla Speedtests conducted in low and high income block groups. The $H_0$ is the number of Speedtests (pp) in low income block groups is not less than that of high income block groups in a given state. We then pose $H$ as the number of Speedtests(pp)in a high income block group is greater than the number of tests that originate from the low income block groups. Results show a large number of states reject the null hypothesis across all quarters ($45$, $48$, $48$, $46$, $45$, and $47$ states in Q1-Q4 of 2020 and Q1-Q2 of 2021, respectively). This indicates that users from high income block groups tend to conduct a greater number of Speedtest  measurements compared to the low income block groups.

\begin{table*}[t]
    \resizebox{0.8\linewidth}{!}{%
  \begin{tabular}{l|cc|cc|cc|cc|cc|cc}
    \toprule
    \multirow{2}{*}{Metric}&
		\multicolumn{2}{c|}{Q1-2020}  & 
		\multicolumn{2}{c|}{Q2-2020} &
		\multicolumn{2}{c|}{Q3-2020} &
		\multicolumn{2}{c|}{Q4-2020} &
		\multicolumn{2}{c|}{Q1-2021} &
		\multicolumn{2}{c}{Q2-2021} \\
		& {K-S} & {M-W U} & {K-S} & {M-W U} & {K-S} & {M-W U} & {K-S} & {M-W U} & {K-S} & {M-W U} & {K-S} & {M-W U}\\
		\midrule
		Download (Mbps) & 21 & 30 & 19 & 28 & 20 & 30  & 27 & 31 & 25 & 32 & 24 & 32\\
		Upload (Mbps) & 31 & 35 & 31 & 35 & 33 & 36 & 32 & 37 & 33 & 36 & 30 & 36 \\
		Latency (ms) & 22 & 27 & 21 & 25 & 21 & 29 & 22 & 26 & 22 & 26 & 21 & 21 \\
    \bottomrule
  \end{tabular}
  }
  \caption{Number of states that strictly conformed with our income hypothesis for all network metrics.}
  \label{tbl_test_results}
\vspace{-20pt}
\end{table*}

\noindent
{\bf Takeaways.} 
Our analysis of the relationship between income and Internet performance produces some key results. First, there  exists multiple states, such as Georgia and Louisiana, where we detect statistically better Internet performance in favor of Speedtest users from high income block groups. This points towards a likely gap in Internet access quality between these two types of income areas in these states. In a parallel work, we have analyzed the pricing structure for Internet services offered by the major Internet service providers around the country. Our preliminary results show that cost of Internet access remains largely invariant of location and income across the country. As a result, the higher tiers of Internet service likely remain out of reach for  lower income populations. On the other hand, states such as New Jersey and Rhode Island do not reveal a relationship between income and Internet quality during our study period. This likely indicates  the pervasiveness of quality Internet access across these states. Finally, our analysis on the number of Speedtests demonstrates one source of bias that exists in crowdsourced  active measurement platforms where the Internet performance of lower income users may be under-represented.

\section{Analysis of States demonstrating Digital Inequality}
\label{sec_deep_dive}

As discovered in Section~\ref{sec_factors}, a gap in  Internet quality is present for users of Speedtest across many states in the dimensions of location and income. Both the K-S and M-W U tests revealed that a vast majority of states demonstrate this divide across urban and rural locations. However, in terms of income, $23$ states did not reveal the presence of digital inequality in any of the six quarters analysed. In this section, we specifically examine the characteristics of the states that conformed with our income hypothesis to determine whether and how they differ from those of the non-conforming states.

\subsection{Area}

Figure~\ref{fig_state_characteristics}(a) shows the distribution of the geographic area of the conforming and non-conforming states. The largest conforming state is Alaska, with an area of roughly $700$K square miles (mi$^2$). Other larger conforming states include Texas and California. On the other hand, the largest non-conforming state (states that failed to conform in any of the six quarters) is Montana, with an area of $150$K mi$^2$. Excluding Alaska, Texas and California, the average size of a conforming state is $60$K mi$^2$; when including these three large states, the average size jumps to $93$K mi$^2$. In both cases, this is larger than the average size of all  non-conforming states ($46$K mi$^2$).  Smaller state size could potentially ease the challenge and cost of network infrastructure deployment, and subsequently make it easier to provide higher Internet quality  
to all populations within a state.


\subsection{Population}

The population distributions of conforming and non-conforming states are demonstrated in Figure~\ref{fig_state_characteristics}(b). California, with a population of $40$M, is the most populous conforming state. In the non-conforming category, Ohio possesses the largest population of $12$M. On average, non-conforming states show much lower populations  ($3.2$M) compared to  conforming states ($8.2$M). The greater populations of the conforming states, coupled with greater geographic size, could cause challenges in network infrastructure deployment, resulting in  disparities in physical equipment location and subsequent  inequality of access across different population groups.

\subsection{Income Dispersion}

\begin{figure*}[t]
\vspace{-20pt}
\centering
\subfloat[Size]{\label{a}\includegraphics[width=.28\linewidth]{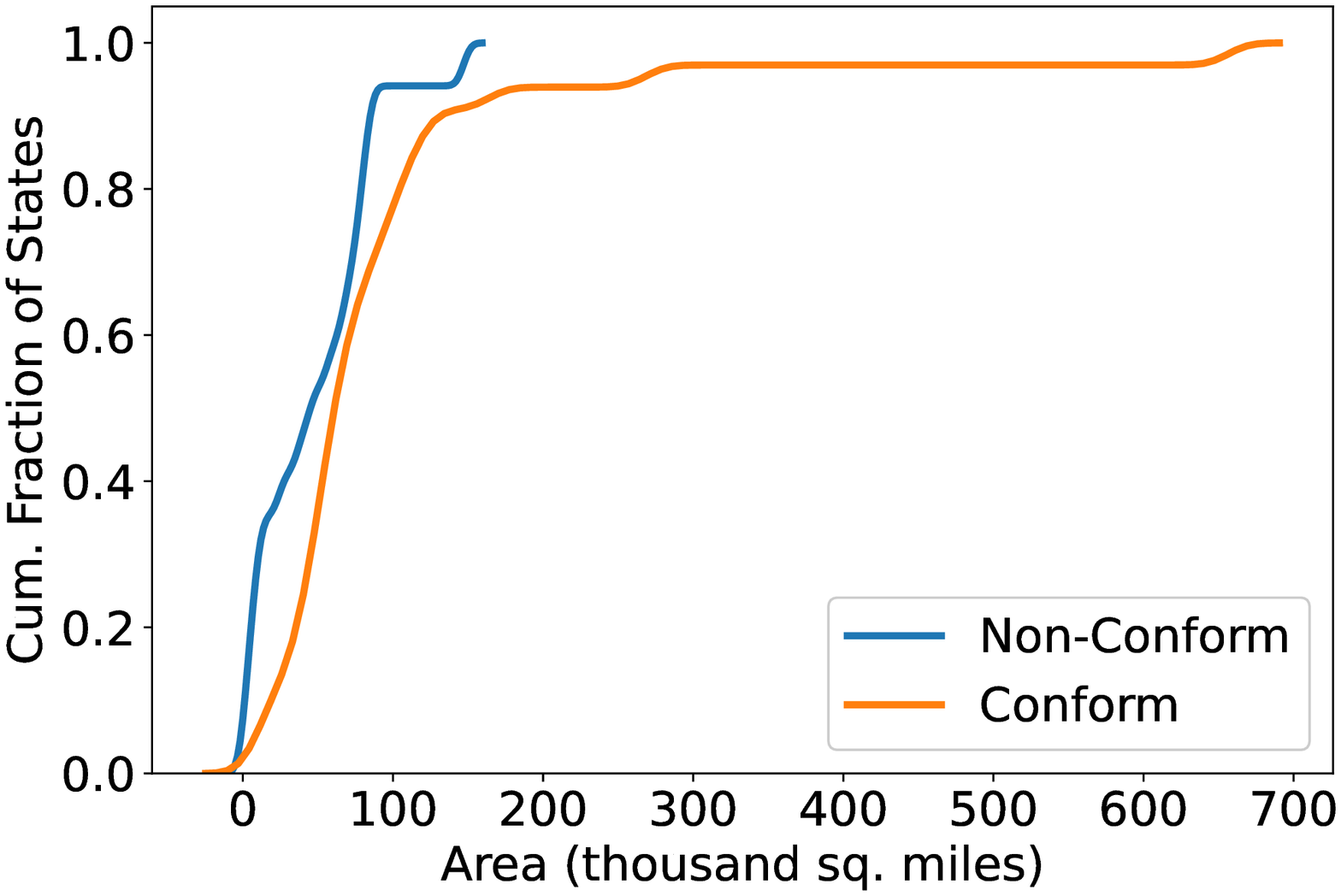}}\hfill
\subfloat[Population]{\label{b}\includegraphics[width=.28\linewidth]{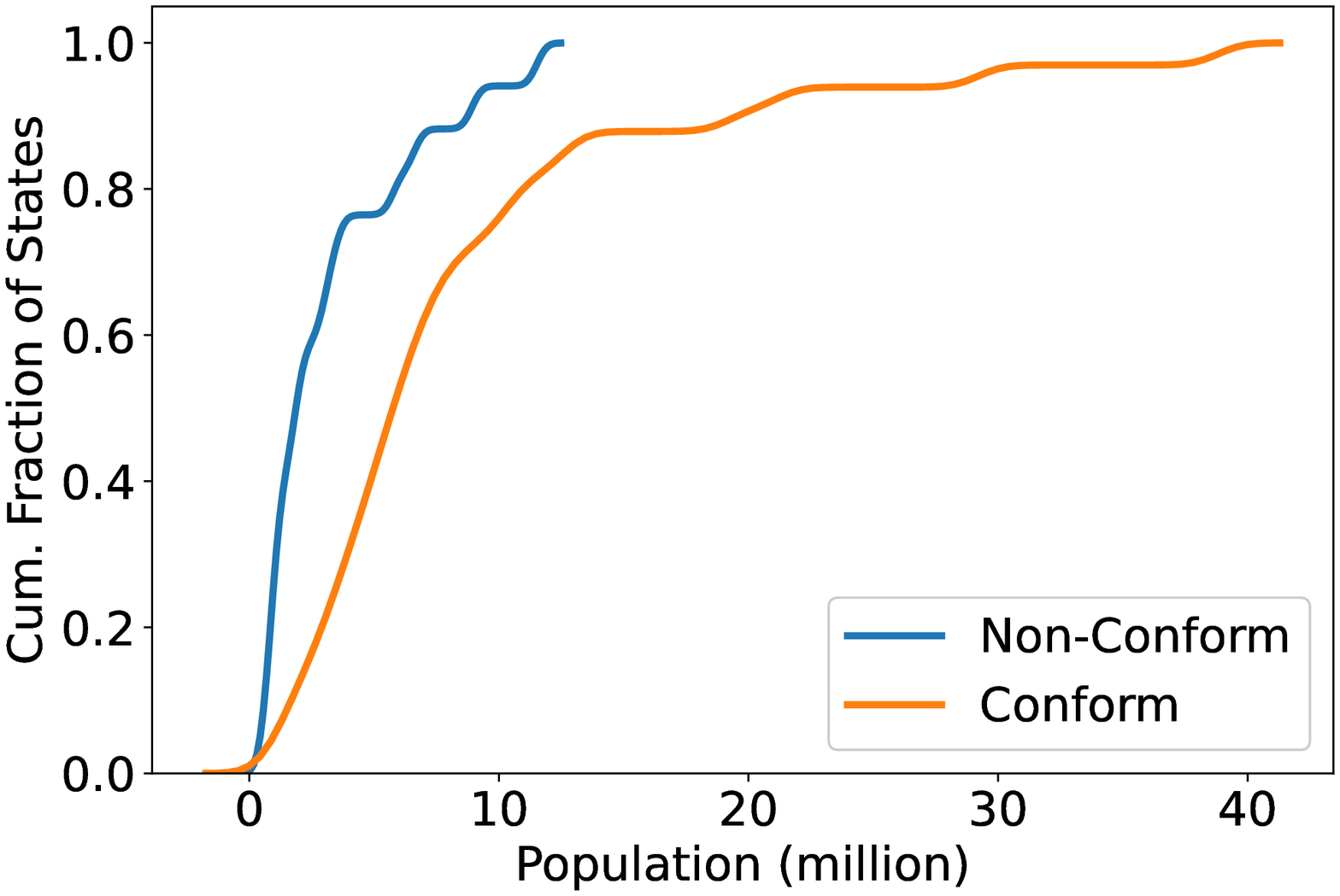}}\hfill
\subfloat[Income Dispersion]{\label{c}\includegraphics[width=.28\linewidth]{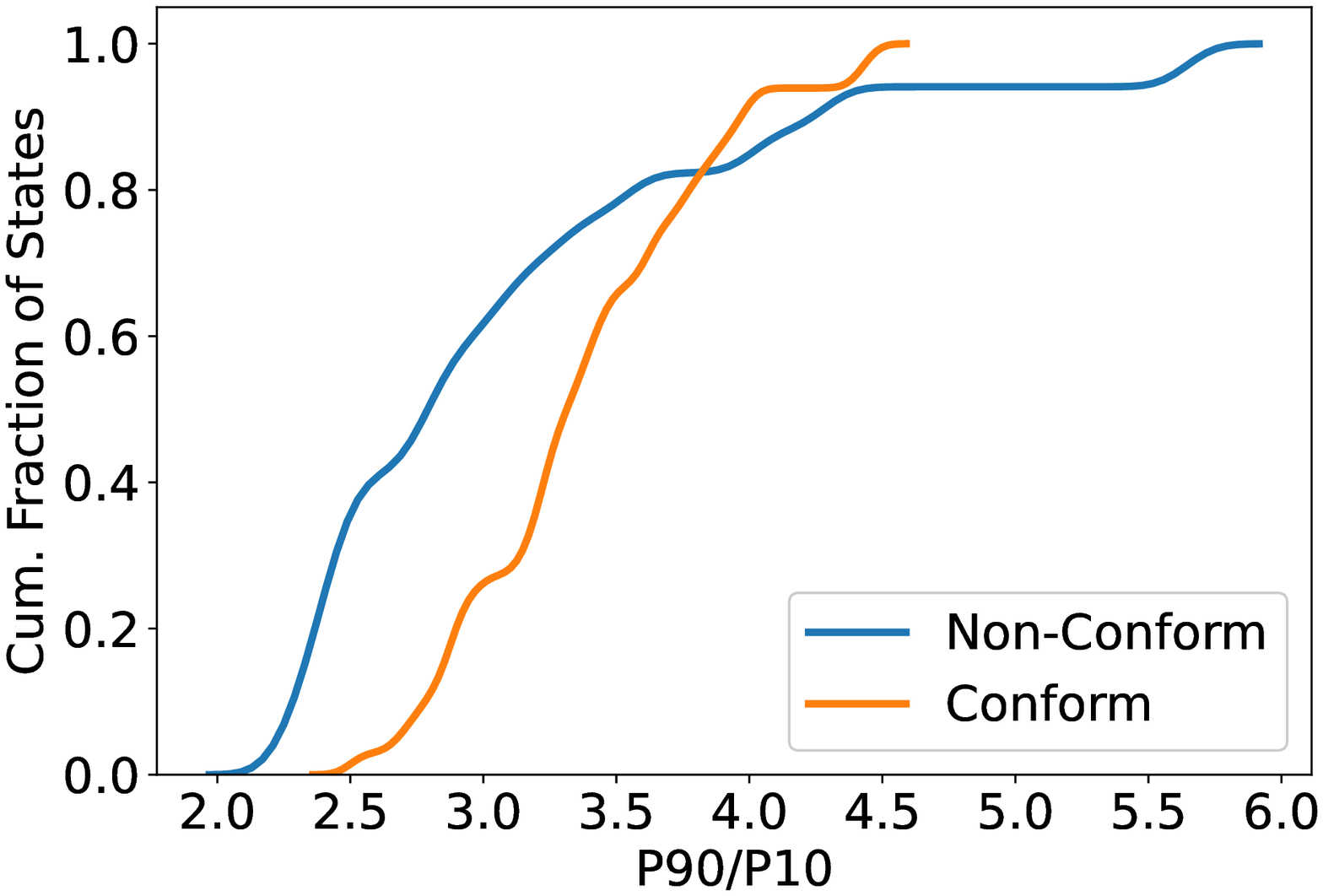}}\hfill

\vspace{-8pt}
\caption{Characteristic distributions  for states with an Internet quality gap between  high  and low income populations.}
\label{fig_state_characteristics}
\vspace{-12pt}
\end{figure*}

To represent the income inequality within a given state, we compute the ratio of the $90^{th}$ percentile (P90) and $10^{th}$ percentile (P10) block group level median household income of each state. A higher $P90/P10$ ratio  indicates a higher dispersion of income within the state. As can seen from Figure~\ref{fig_state_characteristics}(c), conforming states tend to have higher ratio compared to  non-conforming states. This is further illustrated by the higher average ratio of $3.4$ for conforming states compared to $3$ for the non-conforming states. States with higher income dispersion could potentially have a gap in purchasing power that can impact a low income subscriber's ability to purchase higher cost subscription plans, which are typically associated with better Internet quality (download and upload speeds).  
\section{Discussion}
\label{discussion}

In this section, we discuss the significant challenges  associated with  research in the general space of digital inequality. Based on the experience of aggregating the presented data, we also provide recommendations that we hope could lead to additional research in assessing and bridging  digital inequity. 
\subsection{Challenges}
\subsubsection{Lack of Granular Internet Measurement Data} 

Many publicly available Internet measurement datasets, such as the FCC's Measuring Broadband America (MBA) project~\cite{fcc_mba},  lack both expansive and fine-grained geographic coverage. This is in part because of the difficulty in  collecting access measurement data with strong spatial and temporal fidelity due to challenges of  privacy-preserving data collection from user homes. This lack of fine-grained, spatially-diverse Internet measurement data across varied demographic variables presents significant challenges to detailed analysis of Internet quality and affordability~\cite{mapping_Bronzino}. The magnitude of this challenge is highlighted by the recent FCC initiative~\cite{fcc_mobile_challenge}. Through this initiative, the FCC has asked researchers and stakeholders to propose methodologies and techniques to gather high-fidelity, fine-grained data to create more comprehensive and accurate maps of Internet availability and quality.

\subsubsection{Lack of Understanding User Context} A dataset such as the one from Ookla provides information on network quality of service from different vantage points. However, the metrics collected during these tests do not shed light on critical user related information such as the subscribed tier of service and the actual quality of experience for different application genres. Without knowing the ISP and tier of subscription, it is difficult to understand the fundamental reasons behind poor quality of service. Similarly, without information about user quality of experience, it is difficult to determine the usability of different applications. 
Collection of this data in a secure and privacy preserving manner remains an open research problem.

\subsubsection{Lack of Broadband Pricing Data} Through the FCC's Urban Rate Survey~\cite{fcc_urban}, aggregated information related to the price and speed offered by  ISPs is reported at the granularity of county, not individual homes. The U.S. broadband industry suffers from a severe shortage of publicly available datasets that contain information about Internet access plan speeds and pricing. Practices employed by ISPs are difficult to study and analyse in the absence of such information. We hope that our work draws further attention to this research space so that issues related to broadband availability and cost, that in turn adversely affect the penetration of high speed Internet connectivity, can be more deeply studied.

\subsection{Recommendations}

\subsubsection{Publicly Available Data} 

Given the complex nature of digital inequality, the integration of different data types is needed to better characterize and ameliorate its manifestations. 
However, the overhead associated with some data collection efforts can be significant. For instance, the FCC's Measuring Broadband America (MBA) project~\cite{fcc_mba} requires measurement-capable routers to be shipped to volunteers.  Hence it may make sense of incentivize or mandate ISPs to periodically report access quality measurements to/from their subscribers.
Placement of such measurement datasets in the public sphere would significantly 
aid  research that characterizes and pinpoints the specific locations of digital inequities, particularly across diverse user groups.

\subsubsection{Examination of ISP Practices} While our work did not study the Internet service pricing structure and its impact on different population groups, prior work~\cite{ssrn_berkley} has indicated the presence of certain ISP monopolies across different areas of the U.S. Due to the market monopoly,  ISPs could potentially exert Internet pricing that leaves certain customer groups paying for more than what they would otherwise in a market with multiple competing ISPs. These findings point towards a need to conduct an in-depth and extensive examination of ISP competition across states. Careful analysis is also needed to better understand Internet access pricing structures and the role the cost of access plays in digital inequality.

\subsubsection{Adjusted Cost of Internet Access} Our results indicate that Speedtest performance in  lower income areas lags behind that of higher income areas. One potential cause could be the cost of access to high quality Internet service. 
Pricing structures that do not vary based on median income
can have the effect of marginalizing some communities and reducing their ability to  access  higher tiers of Internet service. The Emergency Broadband Benefit initiative by the FCC~\cite{fcc_nytimes} subsidises the cost of high speed Internet for low income households and highlights the need to support certain communities and individuals in their ability to purchase high quality Internet. While an extremely positive step, there have been indications that some ISPs may be forcing customers into more expensive plans in order to take advantage of these subsidies~\cite{ars_verizon,ebb_issue}.  As subsequent assistance plans are rolled out, it is  important to monitor  usage and impact on the populations they are  meant to assist.


\section{Related Work}

Every year, the Census, through the American Community Survey (ACS) One Year estimates, compiles a list of cities with the worst Internet connectivity in the country~\cite{acs:one}. However, this estimate is only done for cities with population greater than $65,000$, leaving other regions unassessed. Critically, it is these smaller communities that are more likely to have sub-par Internet access.  Similar to our work,~\cite{fastly:speed} analysed the relationship between income and download speed at the geographic  granularity of zip codes in the U.S. The work utilized income data (grouped into five income bins) obtained from 2017 tax returns filed with the Internal Revenue Service. The study demonstrated a positive correlation between zip code income and download speed. In~\cite{mapping_Bronzino}, the authors conducted an analysis similar  to ours using the Ookla Open data~\cite{ook_open_data} and demonstrated the variability of important Internet quality metrics  between  communities. Our work goes a step further in that we conduct a comprehensive analysis at finer geographic granularity to understand several dimensions (location, income and cost of access) of Internet access variability.

The work conducted in~\cite{bat} demonstrated that the FCC significantly overestimates coverage and highlighted the  lack of coverage in rural and marginalized communities. The work in~\cite{BischofCharacterizing} showed moderate correlation between  reliability of Internet service (packet loss) and type of area (urban/rural). Through our analysis, we show that the quality of Internet between different states and  different communities within these states also varies.
Other studies have  shown the shortcomings of FCC's Form 477~\cite{fcc_477}. In a recent study conducted by Microsoft~\cite{microsoft-access}, it was estimated that $162.8$ million Americans did not have access to high-speed broadband, a number far greater than the FCC's estimate. A similar study~\cite{bbnow_fcc}
estimated $42$ million ($6.5\%$ more than FCC estimates) Americans  do not have access to broadband Internet. In~\cite{rural-broadband,nexus,minority,income}, demographic factors such as location, race and/or income are all shown to impact Internet access. We advance this body of work and demonstrate that while areas may have Internet access, the quality of that access may differ widely by location and income.

Similar to our work,  the authors of~\cite{igor_ookla} used crowdsourced measurements to benchmark Internet performance across multiple metropolitan areas. In~\cite{marcel-dsl}, cable and DSL  performance in residential areas of North America and Europe was characterized.  In~\cite{TrevisanFive, botFacebook, LiDeployed, KumarThings}, additional work related to understanding Internet performance of different user groups was conducted.  While relevant, these prior studies  did not attempt to understand how the Internet performance varies between users of different locations (urban/rural) or income levels. Finally,  cost effective deployment solutions were proposed to increase coverage in unserved or under-served areas in~\cite{gaia:techno,colte, nsdi_manager}.

\vspace{-8pt}
\section{Conclusion}
\label{conclusion}
Internet inequality continues to persist across the U.S.
Our work integrates
data on Internet performance with location and income to explore multiple dimensions in which this divide manifests amongst users of the popular Ookla Speedtest application. 
Our findings point towards the need to develop more accurate Internet coverage, affordability and quality measurement tools to facilitate more fine-grained analysis of 
the quality of experience of  user groups. Additionally, given the lack of information that currently exists in the broadband market, our work highlights the need for increasing visibility in this segment to better understand the root causes behind Internet access quality differentials between users. 
It is our hope that our findings can help guide the efforts of policymakers and researchers in narrowing this persistent digital gap.
\bibliographystyle{ACM-Reference-Format}
\bibliography{reference}

\end{document}